\documentstyle[preprint,eqsecnum,aps]{revtex}
\begin{document}

\def\A{A}
\def\H{H}
\def\D{D}
\def\m{m}
\def\p{C}
\def\ME{MaxEnt }
\def\e{\eta}
\def\prior{{\it Prior }}
\def\posterior{{\it posterior }}
\def\likelihood{{\it Likelihood }}
\def\crn{\nonumber\\}
\newcommand{\Eq}[1]{Eq.\ \ref{#1}}

\draft
\preprint{}
\title{Consistent Application of Maximum Entropy to\\
Quantum-Monte-Carlo Data}
\author{
W.\ von der Linden$^a$,
R.\ Preuss$^b$,
and W.\ Hanke$^b$\\}
\address{
a)
Max-Planck-Institut f\"ur Plasmaphysik, EURATOM Association\\
D-85740 Garching b. M\"unchen, Germany\\
e-mail: wvl@ibmop5.ipp-garching.mpg.de\\
b) Physikalisches Institut, Universit\"at W\"urzburg, Am Hubland\\
D-97074 W\"urzburg, Federal Republic of Germany\\
e-mail: roland@physik.uni-wuerzburg.de
}

\date{\today}
\maketitle
\begin{abstract}
Bayesian statistics in the frame of the maximum entropy concept has
widely been used for inferential problems, particularly, to infer
dynamic properties of strongly correlated fermion systems from
Quantum-Monte-Carlo (QMC) imaginary time data. In current applications,
however, a consistent treatment of the error-covariance of the QMC data
is missing.  Here we present a closed Bayesian approach to account
consistently for the QMC-data.
\end{abstract}

\narrowtext
\section{Introduction}
\par

Bayesian statistics\cite{Kendall94} provides a general and consistent
frame for logical inference based on incomplete and noisy data and
prior knowledge. Combined with the entropic \prior it is referred to as
quantified maximum entropy \cite{Skilling,Gull} (\ME)
and yields the most probable and unbiased result, given
the data-constraints and prior knowledge.
\ME has originally been introduced \cite{Skilling,Gull}
to infer celestial images from incomplete and noisy
radio-astronomic data. In the sequel, it
has been applied successfully to various other
data-analysis problems\cite{Buck,silv,Wvdl93,Wvdl94}.

Here we will focus on the ill-posed inversion problem encountered
in Quantum-Monte-Carlo (QMC) simulations \cite{silv,sil2}.
In this field, \ME has become a standard and successful technique to
infer dynamic properties of strongly correlated fermion systems from
imaginary time QMC data, which intrinsically is an inverse Laplace
transform.
QMC yields values for dynamic quantities along the
imaginary time-axis for a finite number of times.
The inversion is not unique due to the limited number of
data and the presence of statistical errors.
A direct inversion of the Laplace transform would tremendously overfit
the noise and the desired signal would be buried underneath it.
Bayesian probability theory provides a consistent frame to
separate the signal from the noise.

A complication arises, however, in the present application as the
errors of the QMC-data are correlated.
It has been proposed\cite{silv} to include the error-covariance
matrix as exact data-constraints. This approach has been heavily
debated as it leads to the following dilemma.
For a standard QMC sample size, the off-diagonal
elements of the covariance matrix are not negligible and ought
to be taken into account. However,
the errors of the covariance matrix are huge and the information
provided by the QMC covariance matrix is useless and in many
applications
even disadvantageous. The error of the covariance matrix
can be decreased by increasing the sample size. This procedure
is, however, computationally very expensive and, moreover, needless as
the correlation of the errors decreases at the same time.
The dilemma obviously arises due to the neglect of the errors
of the covariance matrix.
Here we present a closed Bayesian approach to account fully for
the noisy QMC data plus covariance matrix. This scheme allows
to infer reliable results with the least amount of computer
time.

Bayes' theorem aims at determining the posterior probability
(\posterior) $P(\A|\D \p \H)$ of the sought-for quantity $\A$,
given hypotheses $\H$ and QMC data $\D$, which are related to
$\A$ via
\begin{equation}
D^{ex}_l = D_l^{ex}(\{A\})
\label{dex}
\qquad.
\end{equation}
The experimental data $\D_l$ deviate from the exact values by the
statistical QMC errors $\e_l$.
Due to the QMC-algorithm, the errors are correlated and
the information about the error-covariance matrix will be denoted
by $\p$.
Bayes' theorem relates the \posterior to the likelihood
function $P(\D|\A \p \H)$, which contains the error statistic of the
QMC data, and the prior probability $P(\A|\p \H)$
\begin{equation}
P(\A|\D \p \H) =
\frac
  {P(\D|\A \p \H) P(\A|\p \H)}
  {P(\D|\p \H)}
\label{bayes}
\qquad.
\end{equation}
The most honest \prior should
summarize all our prior knowledge --
the knowledge we have about $\A$ prior to receiving the
experimental data $\A$ --
and nothing more, i.e.\ it should be as ignorant as possible otherwise
\cite{Kendall94}.
Our prior knowledge is part of the hypotheses $\H$.
If nothing is known about the solution $\A$ the most ignorant
\prior would simply be $P(\A|\p \H)=const$.
(If nothing is known about a coin which we are going to toss,
common sense tells us to assign equal prior probabilities
to the events ''head (tail) shows up'').

In the case of a positive, additive distribution function (PAD)
-- e.g. the spectral density --
the most ignorant \prior is the entropic \prior
\cite{Skilling_Buck}
\begin{equation}
P(\A|\p \H) = \frac{1}{Z_S}
\exp\Bigl(
   \alpha \underbrace{
                       \int [ \A(\omega) - \m(\omega)
                       - \A(\omega) \ln(
                                        \frac{\A(\omega)}
                                             {\m(\omega)}
                                       ) ] d\omega
                     }_S
    \Bigr)
\qquad.
\label{entropy}
\end{equation}
$Z_S$ is the normalization constant guarantying
$\int  P(\A|\p \H) {\cal D}\A = 1$.
In the following we will assume, as indicated in \Eq{entropy}
that $\A=\A(\omega)$ is a function of the frequency
$\omega$.
The entropy $S$ is measured relative to a default-model $m(\omega)$
which contains ''weak prior assumptions'' which can still be overruled
by the data constraints.

Next we will turn to the determination of the {\it Likelihood},
the central topic of this paper.
Usually this part is dealt with in one sentence:
``the \likelihood function is given by a Gaussian''.
Here we will be more specific why and when this is true.
The \likelihood quantifies the probability for
the realization of the specific data-values $\D$ measured in the
experiment, supposing the exact function $\A$ were known.
Given $\A$, the exact values $\D^{\rm ex}$ in
data-space are also known (\Eq{dex}).
The \likelihood describes therefore the error statistics of the
QMC-data
\begin{equation}
P(\D|\A \p \H) =: \rho(\D^{\rm ex} - \D) = \rho(\e)
\quad{\rm given\ \p\ and\ \H}\quad.
\end{equation}
In the following, \p\ stands for
the error-covariance matrix $C_{ij}$
\begin{equation}
C_{ij} = \frac {1}{N-1} \langle  \e_i \e_j\rangle = \frac {1}{N-1}
\langle (D_i^{\rm ex} -D_i)
(D_j^{\rm ex} -D_j)\rangle
\qquad,
\end{equation}
measured by QMC.
$N$ is the number of data-values $\D_i$.

To begin with, we assume that the exact values of
the error-covariance are known. The resulting problem is to determine
the PAD $\rho(\e)$ given the constraints
\begin{eqnarray}
C_{ij} &=& \int \rho(\e) \e_i \e_j d\e \\
1&=&\int \rho(\e) d\e
\qquad.
\end{eqnarray}

This is a problem falling into the realm of Jaynes' MaxEnt
\cite{Kapur}, which is analogous to deriving the barometric
formula, Maxwell's velocity distribution, or  Fermi- and Bose statistics.
In this framework, $\rho(\e)$ is obtained upon maximizing
the entropy, subject to the exact data-constraints.
Treating the constraints with Lagrange parameters we are out for
the maximum of
\begin{equation}
{\cal L} = S -
\sum_{ij} \lambda_{ij} \bigl( \int \rho(\e) \e_i \e_j d\e -
C_{ij}\bigl) -\lambda_0 \sum_i \bigl( \int \rho(\e) d\e - 1\bigl)
\qquad,
\end{equation}
with $\lambda_0,\lambda_{ij}$ being Lagrange parameters.
Upon maximizing ${\cal L}$ with respect to $\rho(\e)$ one obtains
an analytic expression for the solution
\begin{equation}
\rho(\e) = \frac{1}{Z} e^{-\sum_{ij} \lambda_{ij} \e_i \e_j}
\label{solution}
\qquad.
\end{equation}
An ignorant, flat default model ($\int\m(\omega)d\omega=1$) has been
assumed.
$Z$ is determined via the normalization constraint
\begin{equation}
Z = \int e^{-\sum_{ij} \lambda_{ij} \e_i \e_j} d^N\e
= \frac{\pi^{N/2}}{\sqrt{\det(\lambda_{ij})}}
\end{equation}
The covariance constraint implies
\begin{equation}
C_{ij} = - \frac{\partial \ln(Z)}{\partial \lambda_{ij}}
= \frac{1}{2} (\lambda^{-1})_{ij}
\quad\Rightarrow
\lambda_{ij} = \frac{1}{2} (C^{-1})_{ij}
\label{exact}
\qquad.
\end{equation}
Hence the  \likelihood is the ubiquitous normal distribution
\begin{equation}
\rho(\e) = \frac{1}{\sqrt{\det( 2 \pi C)}}
e^{-\frac12 \sum_{ij} \e_i C^{-1}_{ij} \e_j}
\label{inverse}
\qquad
\end{equation}
which simplifies to a Gaussian if the errors are uncorrelated
$C_{ij} = \delta_{ij} \sigma_i^{2}$
\begin{equation}
\rho(\e) = \frac{1}{\sqrt{\prod( 2 \pi \sigma_i^2)}}
e^{-\frac12 \sum_i \frac{\e_i^2}{\sigma_i^2}}
\qquad
\end{equation}

Unfortunately, this handy result is only valid if the QMC
error-covariance were known exactly, which is not the case.
Therefore, the errors of the covariance matrix have to be treated on
the same footing as the errors of the QMC data \D\ in the first place
-- by quantified MaxEnt\cite{Skilling}.

Again, the \posterior for $\rho(\e)$, given the QMC error-covariance
$C_{ij}$, and the statistical errors $\sigma_{ij}$ of $C_{ij}$ and
all our hypotheses can be determined via Bayes' theorem
\begin{equation}
P(\rho |C \sigma \H) =
\frac
  {P(C|\rho \sigma \H) P(\rho|\H)}
  {P(C|\H)}
\qquad.
\end{equation}
Superfluous conditions have been discarded.
Again the entropic \prior is invoked.
QMC simulations provide the statistical error $\sigma_{ij}$ of the
covariance matrix. Further information is not available by
present QMC simulations. We therefore assume
that the error $\sigma_{ij}$ are uncorrelated and known, this
is part of our hypotheses H.
A generalization beyond this assumption is straightforward.
Along the lines presented above the \likelihood reads
\begin{equation}
P(C|\rho \sigma \H) =
\exp\Bigl(
          -\frac12
             \underbrace{\sum_{ij}
              \frac{(C_{ij}-\int \rho(\e) \e_i \e_j d\e)^2}
                   {\sigma_{ij}^2}}_{\chi^2}
    \Bigr)
\qquad.
\end{equation}

The MaxEnt result is hence obtained upon maximizing
the \posterior
$P(\rho |C \sigma \H) \propto \exp\bigl( \alpha S - \frac12
\chi^2\bigr)$ or rather
\begin{equation}
{\cal L}(\rho,C)  = \alpha S - \frac12 \chi^2
\end{equation}
The solution can be cast into the same functional form (\Eq{solution})
as in the case of exact data-constraints\cite{Silver93,Wvdl94},
merely the determination of the Lagrange parameters
is modified due to the presence of noise to
\begin{eqnarray}
C_{ij} + \alpha \sigma_{ij}^2 &=& \int \rho(\e) \e_i \e_j d\e\crn
&=& \frac{1}{2} (\lambda^{-1})_{ij}
\label{cov}
\qquad.
\end{eqnarray}
This relation agrees with \Eq{exact} for $\sigma_{ij}=0$.
But in practice the errors of the covariance matrix are
considerable. Particular if only few QMC data are available
the inclusion of the errors of the covariance matrix are
essential to obtain an unbiased estimator for the dynamic
quantities:
The covariance matrix $C_{ij}$ is determined from QMC
data $D_i^{\nu}$ via
$\displaystyle C_{ij} = \frac{1}{N-1}\sum_{\nu=1}^{N}
\Delta D^{\nu}_i \Delta D^{\nu}_j$, where $\nu$ represents
the independent measurements (bins) of $D_i$, as discussed
below.
It is obvious that the rank of $C$ is less or equal to the
number of bins ($N$). Hence, if the number of bins is less
than the dimension of the covariance matrix, the
inverse of $C$, entering \Eq{inverse}, does not exist,
and the regularization term in \ref{cov} is essential (no matter
how small $\sigma_{ij}$ is) to determine $\rho(\eta)$.
\par

Hence the \likelihood $P(\D|\A \p \H)$ remains a normal distribution,
merely the covariance is not the QMC error-covariance, it rather
has to be determined via \Eq{cov}. The regularization parameter
$\alpha$ entering \Eq{cov} can be determined either selfconsistently
upon maximizing the marginal \posterior $P(\alpha| C \sigma \H)$
\cite{Skilling,Gull}, or via the historic condition $\chi^2 = N$.
We employ the historic approach since we know that
the number of good degrees of freedom is small and
both stopping criteria will yield essentially
the same result\cite{Gull}.

\section{Application to the spectral properties of strongly
correlated electrons}
\par

As a typical and topical problem we study the dynamic
properties of the Hubbard model which is presently
the subject of intense analytical and numerical studies.
The detailed understanding of the dynamic properties
of strongly correlated electrons is essential for the theoretical
description of the high temperature superconductors.
The Hubbard model reads
\begin{equation}
H = -t \sum_{<i,j>, \sigma} \left(c^\dagger_{i,\sigma} c_{j,\sigma}
+ h.c. \right) + U \sum_i n_{i \uparrow} n_{i \downarrow}
\end{equation}
with the hopping matrix element $t$ between two adjacent sites.
$c_{i,\sigma}^{(\dagger)}$ destroys (creates) an electron of spin
$\sigma$ on site $i$, $<i,j>$ denotes nearest neighbors, $U$ is a
Coulomb repulsion for two electrons of opposite spin on the same site
and $n_{i, \sigma} = c^\dagger_{i,\sigma} c_{i,\sigma}$.

Unfortunately, dynamic properties cannot be measured directly by
QMC simulations.
Dynamical information is provided by Matsubara Greens functions
$D_{l}=-<T_{\tau}P(\tau_l)Q(0)>$ for
discrete $\tau_l$-values on the imaginary time axis, where
$l=n*\beta/L, \ n=0,...,L$ and $L$ is the number of time slices.
$P$ and $Q$ are operators which define the correlation function.
Here we will consider the one-particle properties of
strongly correlated fermions, and the operators are therefore
$P=c$ and $Q=c^{\dagger}$, respectively.
In order to determine the  spectral density
 $A(\omega)$ for real frequencies $\omega$ the
spectral theorem is applied:
\begin{equation}
D_{l}= - \int A(\omega)
\frac{e^{-\tau_l \omega}}{1+e^{-\beta\omega}} d\omega
\end{equation}
which is, as already stated above, an inverse Laplace transformation
problem and pathologically ill-posed.
\par
Further information on the spectrum is provided by making use of the
lowest order moments of $A(\omega)$,
\begin{equation}
\mu_m=\int \omega^m A(\omega) d\omega,
\end{equation}
which are given by commutation relations, $\mu_m=<[c,H]_m,c^{\dagger}>$,
and are of simple shape for m=1,2 \cite{whi91}.
\par
\section{Results}
In order to compare the QMC/ME-data with exact results, we
consider a chain of $N=12$ sites, which is still
accessible by exact diagonalization (ED) techniques.
The QMC-simulations were done for an inverse temperature of $\beta t=20$
($T=0.05t$), where ground state behavior is achieved for this system size
and a comparison with the $T=0$-ED results is possible.
After obtaining thermal equilibrium, up to 640000 sweeps through the
space-time Ising-fields were performed.
\par
Data from consecutive measurements are highly correlated, even
for the same statistical variable $D_l$.
A study of the skewness (third moment) and the kurtosis
(fourth moment) of the data showed, that to get Gaussian behavior at
least 200 measurements, each separated by 4 sweeps, have to be
accumulated to form one bin.
Then the results of a certain number of bins are used for the inversion
process.
\par
But binning the data does not suffice to get rid of all the
correlations.
Still one has to consider the correlations in imaginary time $\tau$
(i.e.\ between $D_l$ and $D_{l'}$).
In Fig.\ 1 the QMC-result for one single bin is compared to the final
shape of the Greens function for $N^{bin}=800$.
Instead of being distributed 'at random' around the average the data
for this bin
are systematically lower than the average for $\tau < \beta/2$ and
systematically higher for $\tau > \beta/2$.
These correlations may be reduced by forming larger and larger bins
(i.e.\ using more and more computation time).
But it is the aim of this paper to show that this is not a
sensible thing to do and that one can do better by
taking into account the correlations
\cite{Jarrell} and particularly by accounting for the statistical
errors of the covariance matrix.

In the following we will discuss the results of the MaxEnt-procedure
considering three cases: (1) Neglecting any information of the
covariance matrix, (2) using the covariance matrix only, and (3) taking
into account both the covariance and its errors.
To determine the dependence on the number of bins and to give an
quantitative argument for the amount of the computational effort, which
has to be taken, we show spectra resulting from QMC-data for 200, 400
and 800 bins (160000, 320000 and 640000 sweeps, respectively).
The number of time-slices, which corresponds to the dimension
of the covariance matrix, is 160 in the present study.
\par
Starting with 200 bins, which is slightly larger than the number of time
slices,
one can see (first column in Fig.\ 2) that neither
the MaxEnt reconstruction of the plain data (Fig.\ 2a)
nor the additional use of the
covariance matrix (Fig.\ 2d) gives a reliable result. In both
spectra the structures are too pronounced and at the wrong
position. It appears that the results are generally better
if the covariance matrix is not included, since the additional
information is treated as exact data-constraints although it
suffers from pronounced statistical noise.
If, however, the statistical errors of the
covariance matrix are taken into account (Fig.\ 2g) the result
reproduces the ED-result very well. There is a small overestimation of
the spectral weight at $\omega \approx -3$ only.
\par
Increasing the number of bins to 400 gives still an overfitted result
for the first case (Fig.\ 2b).
Taking the covariance matrix into account (Fig.\ 2e) shows a slightly
improved spectrum for $\omega > 0$, but for $\omega \approx -2$ the
spectral weight is suppressed completely.
Again the best spectrum is obtained if the errors of the
covariance matrix are properly accounted for (Fig.\ 2h).
The maximum at $\omega \approx -3$ is damped to the correct shape and
the agreement with the ED-result is nearly perfect now.
\par
Eventually for the large number of 800 bins all three spectra show
satisfactory results (third column of Fig.\ 2). Only in the first case
(Fig.\ 2c) the ME-curve decreases still too fast for $\omega > 5$
leading to a wrong width.
\par
The convergency of the various approaches is reasonable, since
with increasing number of bins, the correlation of the QMC-errors
for different imaginary times vanishes and the covariance matrix
becomes diagonal and the covariance of the errors can be
ignored. At the same time, the errors of the covariance matrix
decrease and assuming exact data-constraints becomes also exact.
\par
Further investigations of the effect of the error of the covariance
matrix revealed that the procedure can be simplified by
assuming a sufficiently large constant relative error
(in our case $20\%$). The results show no significant deviation
from the results obtained by taking the
correctly generated errors.

\section{Conclusion}
We have shown that the imaginary time data, obtained by standard
QMC simulations, suffer from strongly correlated statistical
errors if only a small sample is used. It appears reasonable
to include this additional knowledge (covariance matrix).
But it has been generally observed,
the QMC data of the covariance matrix
are useless if the sample size is small
and they are needless if the sample size is large.
And it seemed generally advantageous to ignore the
off-diagonal elements of the covariance matrix altogether.
At any rate,
this  approach demands extremely long QMC-runs
(more than half a million sweeps was not even sufficient in our case)
and  gets increasingly
impossible for larger system sizes (the computation time for
one sweep scales with $\sim N^3$).
We have shown that only if the covariance matrix and its errors are
treated consistently in the Bayesian frame reliable results can be
obtained regardless of the number bins.

\par
\section{Acknowledgments}
We would like to thank R.\ Silver and S.R.\ White for stimulating
discussions and we
are grateful to the Bavarian "FORSUPRA" program on high $T_c$ research
for financial support.
The calculations were performed at HLRZ J\"ulich and at LRZ M\"unchen.

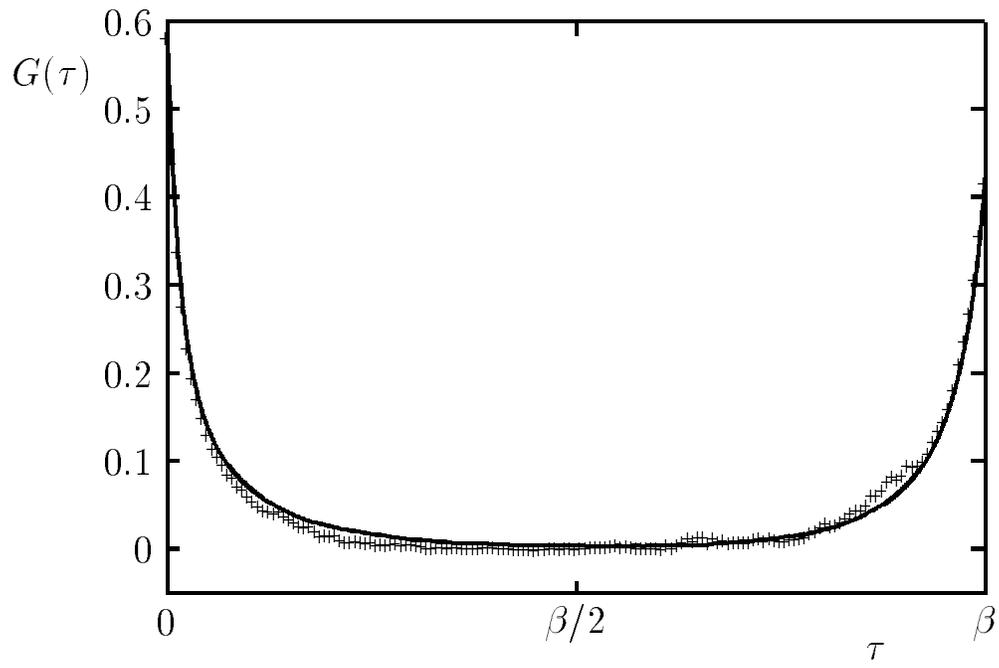
\begin{figure}
\caption{
The data from one bin (points) compared to the final average over 800 bins
(line).
}
\end{figure}

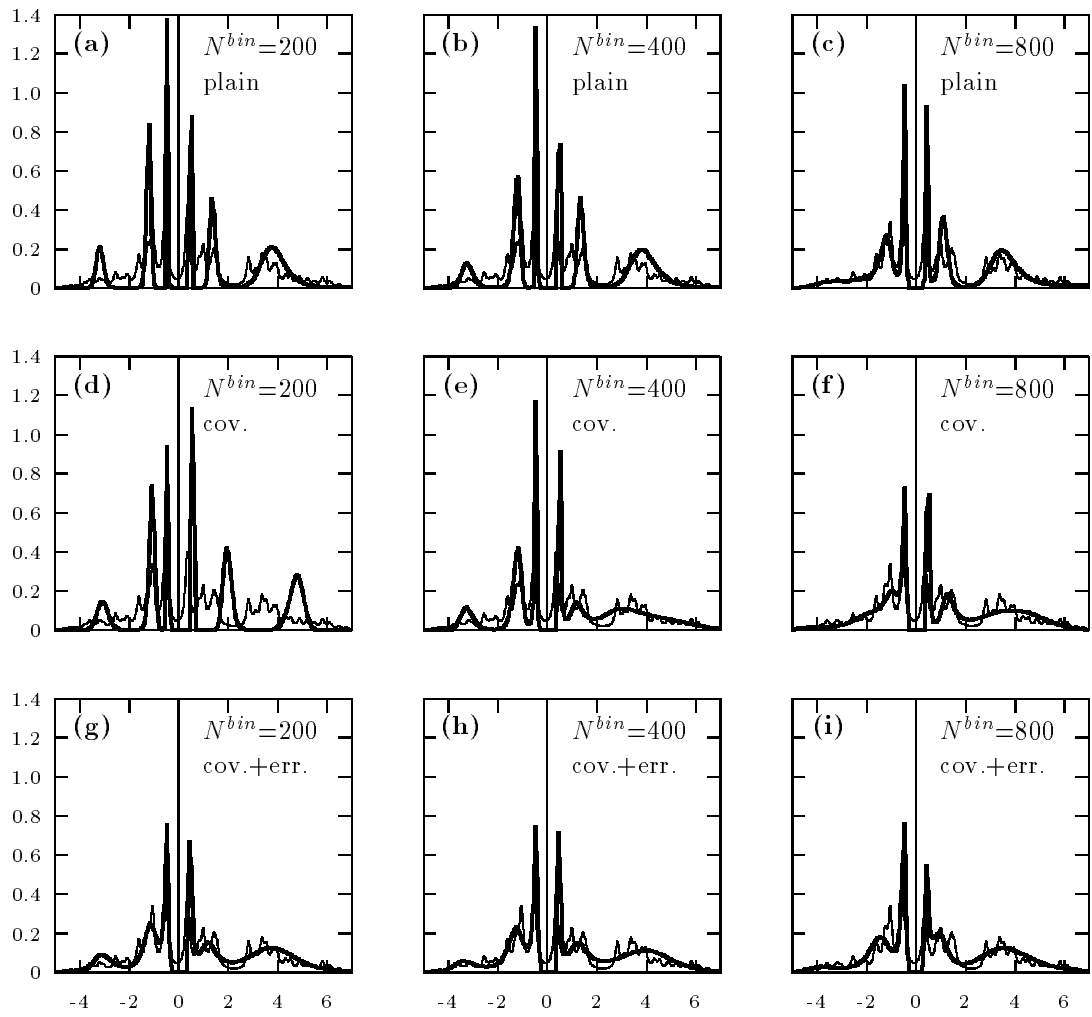
\begin{figure}
\caption{
Comparison of the MaxEnt-spectra (thick line) with the
ED-result (thin line) for different parameters: first row (a,b,c):
without use of the covariance (first case, see text), second row
(d,e,f): with covariance (second case), third row (g,h,i): with the
error of the covariance matrix (third case);
in dependence of the number of bins: first column (a,d,g): 200 bins,
second column (b,e,h): 400 bins, third column (c,f,i): 800 bins.
}
\end{figure}


\begin{references}
\bibitem{Jarrell} %14
             In this section we followed suggestions of a talk
             manuscript by Mark Jarrell, University of Cincinnati.
\end{references}
\end{document}